\documentclass[journal]{IEEEtran}
\usepackage[margin=2cm]{geometry}
\usepackage{aiaimath}
\usepackage{amsmath}
\usepackage{pgf}
\pgfrealjobname{main}
\usepackage{siunitx}
\DeclareSIUnit\photon{photons}
\DeclareSIUnit\pixel{pixel}
\usepackage[nolist]{acronym}
\usepackage{booktabs}
\DeclareMathOperator{\mD}{D}
\DeclareMathOperator{\mR}{R}

\usepackage{lagsim}
\usepackage{motionsimrecons}

\DeclareMathOperator{\mN}{\mathcal{N}}

\begin{document}

\begin{acronym}
	\acro{CBCT}{Cone-Beam CT}
	\acro{MDCT}{Multi-Detector CT}
	\acro{MBIR}{Model-Based Iterative Reconstruction}
	\acro{RMSE}{Root Mean Squared Error}
    \acro{SID}{Source-Isocenter Distance}
    \acro{SDD}{Source-Detector Distance}
\end{acronym}

\title{High-Fidelity Modeling of Detector Lag and Gantry Motion in CT Reconstruction}

\author{{Steven~Tilley~II, Alejandro~Sisniega, Jeffrey~H.~Siewerdsen, J.~Webster~Stayman}%
\thanks{Department of Biomedical Engineering, Johns Hopkins University. email:~web.stayman@jhu.edu.}}

\maketitle
\pagestyle{empty}
\thispagestyle{empty}

\begin{abstract}
    
    Detector lag and gantry motion during x-ray exposure and integration both result in azimuthal blurring in CT
    reconstructions. These effects can degrade image quality both for high-resolution
	features as well as low-contrast details. 
	In this work we consider a forward model for model-based iterative reconstruction (MBIR) that
	is sufficiently general to accommodate both of these physical effects. 
	We integrate this forward model in a penalized, weighted, nonlinear least-square style
	objective function for joint reconstruction and correction of these blur effects.
    We show that modeling detector lag can reduce/remove the characteristic lag artifacts in
    head imaging in both a simulation study and physical experiments. Similarly, we show that
	azimuthal blur ordinarily introduced by gantry motion can be mitigated with proper reconstruction models.
	In particular, we find the largest image quality improvement at the periphery of the field-of-view
	where gantry motion artifacts are most pronounced. These experiments illustrate the generality of the
	underlying forward model, suggesting the potential application in modeling a number of physical effects
	that are traditionally ignored or mitigated through pre-corrections to measurement data.

\end{abstract}

\section{INTRODUCTION}

The need for high-resolution, quantitatively accurate CT reconstructions has
increased with the rise of application-specific systems. 
For example, \ac{CBCT} mammography \cite{Boone2006} and 
extremities systems \cite{marinetto_quantification_2016} require high resolution
to detect microcalcifications and visualize fine trabecular structure,
respectively. Point-of-care \ac{CBCT} head imaging \cite{xu_technical_2016} similarly requires highly accurate reconstruction of
relative attenuation values to detect low contrast bleeds. Such dedicated imaging systems often 
use flat-panel detectors, which are selected for their high-resolution capability and ease of integration into compact systems.
However, a number of physical effects including scintillator blur and detector lag can degrade measurement data,
challenging the above applications. Similar examples of hardware limitations challenging particular applications
can be found in traditional \ac{MDCT}. For example, cardiac and emergency room scanning place high demands on lowering the scan 
time. The high rotation rates in such applications can result in significant blurring effects due to gantry motion
during the integration time of the detector.

Previous work has suggested that such hardware limitations can be compensated through explicit modeling and
incorporation into a \ac{MBIR} algorithm. In particular, we have found that scintillator blur and focal-spot blur in flat-panel systems
can be modeled for potential resolution recovery \cite{TilleyPenalizedLikelihoodReconstructionHighFidelity2017}.
The forward model used in that work is very general and permits incorporation of a wide range
of physical effects. In this work we adopt the same mathematical form for the underlying forward model and apply \ac{MBIR} to
detector lag and gantry motion. 

Detector lag results from the detector trapping and later releasing charge, causing a fraction of the signal from previously acquired projections
to be added (temporally blurred) into subsequent 
projections \cite{AntonukEmpiricalinvestigationsignal1997}, \cite{SiewerdsenEmpiricaltheoreticalinvestigation1997}.
Detector lag effects are usually low contrast and extend across large areas of the reconstruction, originating near 
high contrast objects. A classic example of lag artifacts are the low contrast trails arcing off the skull into the
brain in flat-panel-based head imaging. Traditionally, detector lag corrections are applied through preprocessing 
the measurements prior to reconstruction \cite{Starmannonlinearlagcorrection2012, StarmanInvestigationoptimallinear2011}. To our knowledge, this work
is the first attempt to correct for lag within the forward model of an \ac{MBIR} approach.

Gantry motion blur shares some similarity with lag in that there is an effective blurring over angle. However, this blur occurs within a
single measurement - effectively integrating an arc of projection images based on how far the source and detector have rotated during
an integration period. Such blur exhibits as an azimuthal smearing of the CT volume and is most pronounced toward the edge of the field
of view. Gantry motion effects have been addressed in hardware (e.g., collecting data with a step-and-shoot protocol or more complicated methods \cite{NowakTimedelayedsummationmeans2012})
and in software (e.g., incorporating a blur model into a linearized forward model for \ac{MBIR} \cite{cant_modeling_2015}).

In this paper, we introduce specific models of detector lag and gantry motion, and integrate those models into the general form in 
\cite{TilleyPenalizedLikelihoodReconstructionHighFidelity2017}.
Simulation studies are conducted for both blur scenarios. Image reconstructions are performed using both traditional (unmodeled blur) and
the proposed high-fidelity models, and the resulting images are compared. Preliminary physical-experiment results using a head phantom and a \ac{CBCT}
test bench are also shown to illustrate application in a real system.

\section{METHODS}

Both detector lag and gantry motion blur scenarios can use the same general forward model presented in
\cite{TilleyPenalizedLikelihoodReconstructionHighFidelity2017}:
\begin{subequations}
    \begin{equation}
        \bar{\vect{y}} = \mat{B} \exp\left(-\mat{A} \vect{\mu}\right) \label{eq:fmean}
    \end{equation}
    \begin{equation}
        y \sim \mN\left(\bar{\vect{y}}, \mat{K}\right) \label{eq:fnoise}
    \end{equation}\label{eq:forward}
\end{subequations}
where $\mat{B}$, $\mat{A}$, and $\mat{K}$ are matrices, $\vect{\mu}$ is a vector of attenuation values,
and $\vect{y}$ is a vector of measurements. 
The corresponding penalized likelihood objective function is
\begin{equation}
    \left\| \left( \vect{y} - \mat{B} \exp\left(-\mat{A} \vect{\mu}\right) \right) \right\|^2_{\mat{K}^{-1}} + \beta \mR(\vect{\mu}) \label{eq:obj}
\end{equation}
where R is a penalty function and $\beta$ is the penalty strength.
The $\vect{\mu}$ that minimizes \eqref{eq:obj} is the reconstruction.
A traditional forward model has $\mat{A}$ as the system matrix,
$\mat{B}$ as a diagonal matrix which scales measurements by a gain factor (e.g., photon flux, etc.), 
and $\mat{K}$ as a diagonal matrix of measurement
variances. However, the reconstruction method in \cite{TilleyPenalizedLikelihoodReconstructionHighFidelity2017} which minimizes the objective function \eqref{eq:obj}
makes few assumptions about these matrices, allowing the forward model \eqref{eq:forward} to incorporate many physical properties. 
This reconstruction method may utilize ordered subsets and Nesterov momentum acceleration \cite{Nesterov2005, Kim2015}.

\subsection{Detector lag}

Detector lag may be modeled as a convolution blur where the blur kernel is a sum of exponentials \cite{StarmanInvestigationoptimallinear2011}:
\begin{equation}
    h[k] = \begin{cases} b_0 \delta[k] + \sum_{i=1}^3 b_i \exp\left( - k a_i \right) & \text{if } 0 \le k < K \\
        0 &\text{otherwise}
        \end{cases}.\label{eq:lagker}
\end{equation}
The $K$ parameter in \eqref{eq:lagker} determines the length of the blur kernel (i.e., the number of nonzero terms).  
This convolution is incorporated into $\mat{B}$ in \eqref{eq:fmean}. Specifically, each row of $\mat{B}$ weights and combines a series
of unblurred measurement data to form a measurment with lag.
Physical blur kernel parameters for our test bench system were estimated from the falling edge of a bare-beam scan \cite{StarmanInvestigationoptimallinear2011}.
The estimated parameters used throughout this work are shown in Table~\ref{tab:ker}.
Because $\mat{B}$ is no longer block diagonal with regards to projection number (i.e., $\mat{B}$ blurs among projections),
we cannot trivially apply ordered subsets to speed convergence \cite{NuytsModellingphysicsiterative2013}. 

\begin{table}
    \caption{Blur kernel parameters}\label{tab:ker}
    \centering%
    \begin{tabular}{r  r  r  r  r}
        & 0 & 1 & 2 & 3\\
\midrule
a & --- & 0.998 & 0.0991 & 0.0152\\
b & 0.965 & 0.0165 & 0.000572 & 4.51e-05\\

    \end{tabular}
\end{table}

A simulation study was conducted with an ellipsoidal ``head'' phantom of fat  
surrounded by bone. Data were generated from a phantom with
\SI{0.25x0.25}{\milli\meter} voxels on a system with  %
\SI{580.0}{\milli\meter} \ac{SID} and \SI{800}{\milli\meter} \ac{SDD}. Data
were projected onto a detector with \SI{0.278}{\milli\meter} pixels %
with \SI{0.5e5}{\photon\per\pixel} over \ang{360} in \ang{1} increments.
Poisson noise was added and data were binned by a factor of two, resulting in
\SI{0.556}{\milli\meter} pixels with \SI{e6}{\photon\per\pixel}. We then  %
blurred the data by the calculated blur kernel with a length of $K=359$ and
added readout noise ($\sigma_{ro} =$ \SI{7.12}{\photon}).
Blurring the data after adding Poisson noise correlates the noise as in real systems \cite{SiewerdsenEmpiricaltheoreticalinvestigation1997}.

Data were reconstructed
with \SI{0.5x0.5}{\milli\meter} voxels, a quadratic regularizer, and the separable footprints projector \cite{Long2010}. Two %
reconstruction methods were used: identity blur modeling (i.e., no blur modeling) and detector lag blur modeling (with a kernel length of $K=101$).
In this work we assume uncorrelated 
noise for simplicity, specifically
\begin{equation}
    \mat{K}  = \mD\{\vect{y}\} + \sigma_{ro}^2,\label{eq:uncorr}
\end{equation}
where $\mD\{\cdot\}$ is a diagonal matrix with its argument on the diagonal).
We used \num{5000} iterations and Nesterov acceleration.
Reconstructions were noise matched by varying $\beta$ and taking the standard deviation of the attenuation values
in the center of the image.

Additionally, we scanned a physical head phantom on a \ac{CBCT} test bench with parameters similar to those in the simulation study, 
except projection data were acquired in half angle increments.
In order to focus on only detector lag in this preliminary study, we corrected the data for beam hardening due to water, scatter, and glare,
as described in \cite{0031-9155-60-4-1415}.
Data were reconstructed with \SI{0.5x0.5x0.5}{\milli\meter} voxels using the same blur models as
the simulation study (the blur model used a lag kernel length of $K=201$). Nesterov acceleration was used with \num{4000} iterations. We used a quadratic regularizer, and the same regularization strength
for both reconstructions.

\subsection{Gantry motion}

\begin{figure}
    \leavevmode\beginpgfgraphicnamed{motionsimphantom}%
                        \input{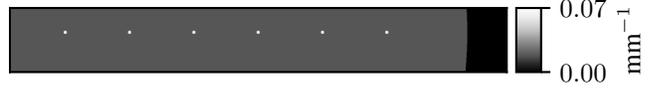}%
                        \endpgfgraphicnamed%
                        
    \caption{A portion of the digital phantom for motion blur studies. The left most circle in this figure is at the center of the phantom. The circles are separated by \SI{20}{\milli\meter}.}\label{fig:motionsimphantom}
\end{figure}

Gantry motion blur is the result of a continuous integration over angle, and may be modeled as
\begin{equation}
    \bar{\vect{y}}_i = \mat{B}_2 \int_{\psi = \theta_i - \Delta \theta / 2}^{\theta_i + \Delta \theta / 2} \exp(-\mat{A}_\psi \vect{\mu}) \mathrm{d}\psi
\end{equation}
where $\bar{\vect{y}}_i$ is the mean measurement vector at projection $i$ and gantry angle $\theta_i$, $\Delta\theta$ is the angular distance over which data is collected for projection $i$, and
$\mat{A}_\psi$ is the projection matrix at angle $\psi$.
A discrete approximation is achieved by oversampling in projection angle and summing the results to obtain the measurement sampling:
\begin{gather}
    \bar{\vect{y}}_i = \mat{B}_2 J^{-1} \sum_{j=0}^{J} \exp\left(-\mat{A}_{\psi} \vect{\mu}\right) \label{eq:motion} \\
    \psi(j) = \theta_i + \Delta \theta \left( j / J - 1 / 2 \right)
\end{gather}
where  $J$ is the angular oversampling factor. $\mat{B}$ from
\eqref{eq:forward} contains $\mat{B}_2$ and the summation term in
\eqref{eq:motion}, and $\mat{A}$ contains all the $\mat{A}_\psi$ used in \eqref{eq:motion}. For example, if the measurement data contains \num{360}
projections and $J$ is 3, then $\mat{A}$ results in \num{1080} projections, and
every three consecutive projections are summed together as part of $\mat{B}$.

\begin{figure}
    \leavevmode\beginpgfgraphicnamed{lagsim}%
                        \input{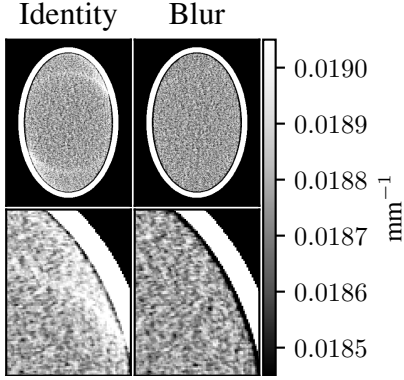}%
                        \endpgfgraphicnamed%
                        
    \caption{Simulation phantom reconstructions with the identity model (left)
    and the detector lag model (right). The second row of images shows a smaller portion of the phantom for better visualization.}\label{fig:lagsim}
\end{figure}

\begin{figure}
    \leavevmode\beginpgfgraphicnamed{lagbench}%
                        \input{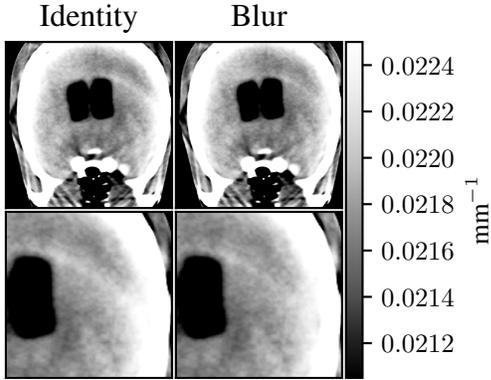}%
                        \endpgfgraphicnamed%
                        
    \caption{Head phantom bench reconstructions with the identity model (left) and the detector lag model (right). The second row shows a smaller portion of the phantom to better visualize the detector lag effect.}\label{fig:lagbench}
\end{figure}

\begin{figure*}
    \centering
    \leavevmode\beginpgfgraphicnamed{motionsimbiasnoise}%
                        \input{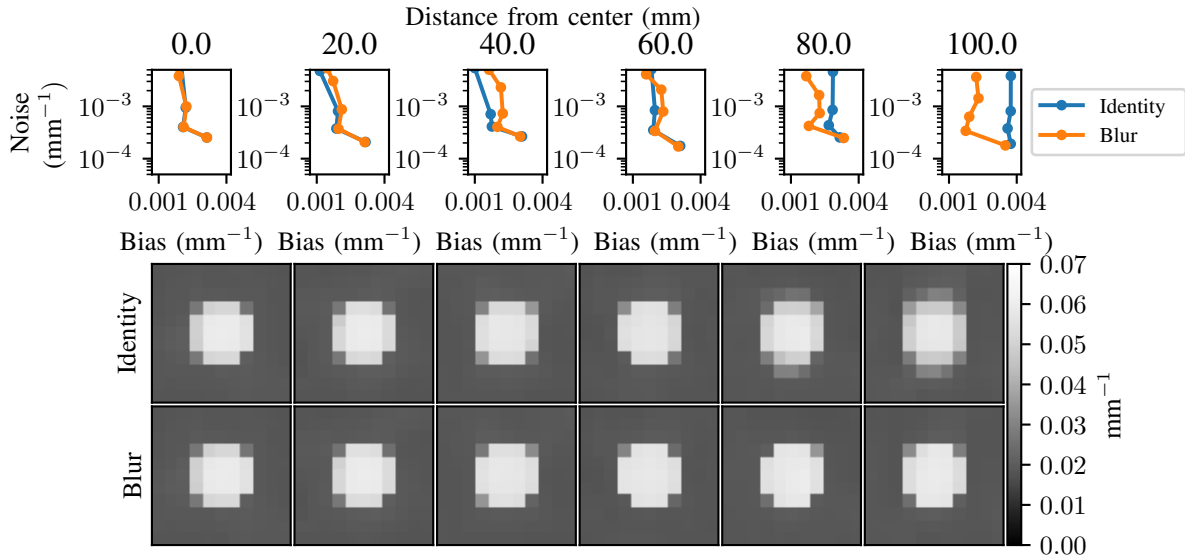}%
                        \endpgfgraphicnamed%
                        
    \caption{Bias/noise curves (top) and reconstructions (bottom) for each ROI in Fig.~\ref{fig:motionsimphantom}. Each column corresponds to a distance from the center of rotation. The top row reconstructions use the identity model and the bottom row reconstructions use the gantry motion blur model.}\label{fig:motionsimbiasnoise}
\end{figure*}

\begin{figure}
    \leavevmode\beginpgfgraphicnamed{motionsimquadcmp}%
                        \input{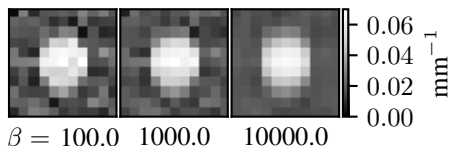}%
                        \endpgfgraphicnamed%
                        
    \caption{Quadratic penalty reconstructions of the \SI{100}{\milli\meter} ROI with blur modeling.}\label{fig:motionsimquadcmp}
\end{figure}

A circular simulation phantom with a diameter of \SI{25}{\centi\meter} and multiple round ROIs at different distances from the center of rotation
was used to evaluate the proposed algorithm.
A subset of this phantom is shown in Fig.~\ref{fig:motionsimphantom}.
A continuous motion system was
simulated with \SI{500}{\milli\meter} \ac{SID}, \SI{1000}{\milli\meter}
\ac{SDD}, and 1000 projections per rotation. This geometry was chosen to approximate high-resolution \ac{MDCT} systems.
Data were generated from a phantom
with \SI{0.1x0.1}{\milli\meter} voxels and a detector with  %
\SI{0.125}{\milli\meter} pixel pitch. We projected at \num{51000} equally %
spaced angles over a \ang{360} rotation. Poisson noise was added prior to binning to
\num{1000} projections and spatially binning to \SI{0.25}{\milli\meter}  %
pixels. The photon flux after binning was \SI{e5}{\photon\per\pixel}. Finally, readout noise was added to the data ($\sigma_{ro} =$ \SI{7.12}{\photon}).

Data were reconstructed with \SI{0.2x0.2}{\milli\meter} voxels.  %
We
used two blur models: an identity blur model (no blur, $\mat{A}$ produces \num{1000}
projections), and a gantry motion blur model with an angular oversampling
factor of $J=5$ ($\mat{A}$ produces 5000 projections). We used an uncorrelated
noise model \eqref{eq:uncorr}, the Huber penalty ($\delta = $ \num{e-3}) \cite{huber_robust_statistics}, and the separable footprints projector \cite{Long2010}. 
Nesterov acceleration was used with \num{1000} iterations and \num{10} subsets.
Bias/noise measurements were
calculated for each ROI. Bias was the \ac{RMSE} between a noiseless
reconstruction and truth at the ROI, and noise was the \ac{RMSE} between a
noisy reconstruction and a noiseless reconstruction in a nearby region.
Bias and noise were calculated for multiple penalty strengths to obtain a bias/noise curve for each method.
Data were also reconstructed with a quadratic penalty and $J=5$ to compare this penalty to the Huber penalty.

\section{RESULTS}

\subsection{Detector lag}

The detector lag digital phantom reconstructions are shown in
Fig.~\ref{fig:lagsim}. The reconstructions are approximately noise matched ---
\SI{\lagsimident}{\per\milli\meter} (identity) and
\SI{\lagsimuncorr}{\per\milli\meter} (blur). When no blur model is
used, detector lag causes a bright trail atrifact arcing off the skull and into the interior of the head. When blur modeling is used, this effect is
eliminated. When lag modeling was applied to bench data, the bright trail off the
skull was dramatically reduced (Fig.~\ref{fig:lagbench}). The fact that the trail was not completely removed may be due to
an insufficient number of iterations (non-converged estimate) or an inaccurate estimate of the lag blur kernel.

\subsection{Gantry motion}

Gantry motion results are summarized in Fig.~\ref{fig:motionsimbiasnoise}.
The bias/noise tradeoff is shown for each ROI at varying distances from the center of rotation. 
The identity model suffers from increased
bias at large distances from the center of rotation, while the blur model bias
is relatively unchanged (suggesting a recovery of spatial resolution). The identity model appears to outperform
the blur model at \SIrange{20}{60}{\milli\meter} from the center of rotation,
although the difference is small. These results are confirmed in
the reconstructions in Fig.~\ref{fig:motionsimbiasnoise}. These reconstructions were approximately noise
matched at the ROI furthest from the center of rotation by altering penalty strength (noise is  
\SI{\motionidentnoise}{\milli\meter} for the identity model and
\SI{\motionblurnoise}{\milli\meter} for the blur model). The circles in the
identity model reconstruction get blurrier along the direction of rotation as
distance from the center increases. However, with the blur model the circles
are accurately reconstructed. Additionally, the blur model's bias
improvement in the \SIrange{20}{60}{\milli\meter} range is difficult to
visualize.

Fig.~\ref{fig:motionsimquadcmp} shows the \SI{100}{\milli\meter} ROI reconstructed with the blur model and the quadratic penalty at three different penalty strengths.
With this penalty the blur model is unable to deblur the circle without a substantial increase in noise.

\section{DISCUSSION}

We have shown that the general reconstruction method presented previously
\cite{TilleyPenalizedLikelihoodReconstructionHighFidelity2017} is capable of
reducing effects due to detector lag and gantry motion blur. The methods
presented here could trivially be extended to model detector lag with other
forms (i.e., not sum of exponentials) or more complicated forms of gantry
motion (e.g., when data acquisition only occurs during a fraction of the
rotation). Additionally, these models may be combined with each other or other
forms of blur, such as focal spot blur and scintillator blur, to further improve image quality.

A major limitation of modeling detector lag is the inability to use ordered
subsets to speed convergence. In practice, one may initialize with a reconstruction without a
lag model and with ordered subsets to get a relatively accurate estimate, and
then reconstruct with the lag model for a handful of iterations. Additionally, a
more accurate initialization may be obtained by lag correcting the projection data
prior to simple \ac{MBIR} (i.e., without lag modeling), and then the final reconstruction obtained with a few iterations
with the full lag model and the original, uncorrected measurement data.

While this work is still preliminary, we note that the edge preserving Huber
penalty plays an important role in the gantry motion reconstructions. We believe
the quadratic penalty's tendency to enforce smooth edges prevents the fidelity  
term from deblurring the gantry motion effects. In contrast, the Huber penalty
doesn't penalize sharp edges to the same degree, and allows the fidelity term
to deblur the image. 
Ongoing work will further explore these issues by analyzing different penalties 
(e.g., sweeping the $\delta$ parameter) and using more complicated image quality targets.

High-fidelity system modeling with \ac{MBIR} can improve image quality by
overcoming hardware limitations such as detector lag and gantry motion.
However, application specific systems may have different limitations and
constraints. The forward model and \ac{MBIR} algorithm used in this work are
sufficiently general to accommodate many physical effects, and may therefore be
used to improve image quality and quantitative accuracy in a wide range of
clinical scenarios.

\section{ACKNOWLEDGMENTS}

This work was supported in part by NIH grant F31~EB023783. The
bench data used in this work was acquired with support of an academic-industry
partnership with Carestream (Rochester, NY). The authors would like to thank
Ali Uneri and Yoshi Otake for GPU software contributions.

\bibliography{ZoteroLibrarybibtex}

\begin{thebibliography}{10}
\providecommand{\url}[1]{#1}
\csname url@samestyle\endcsname
\providecommand{\newblock}{\relax}
\providecommand{\bibinfo}[2]{#2}
\providecommand{\BIBentrySTDinterwordspacing}{\spaceskip=0pt\relax}
\providecommand{\BIBentryALTinterwordstretchfactor}{4}
\providecommand{\BIBentryALTinterwordspacing}{\spaceskip=\fontdimen2\font plus
\BIBentryALTinterwordstretchfactor\fontdimen3\font minus
  \fontdimen4\font\relax}
\providecommand{\BIBforeignlanguage}[2]{{%
\expandafter\ifx\csname l@#1\endcsname\relax
\typeout{** WARNING: IEEEtran.bst: No hyphenation pattern has been}%
\typeout{** loaded for the language `#1'. Using the pattern for}%
\typeout{** the default language instead.}%
\else
\language=\csname l@#1\endcsname
\fi
#2}}
\providecommand{\BIBdecl}{\relax}
\BIBdecl

\bibitem{Boone2006}
J.~M. Boone and K.~K. Lindfors, ``Breast {{CT}}: Potential for breast cancer
  screening and diagnosis.'' \emph{Future oncology (London, England)}, vol.~2,
  no.~3, pp. 351--356, 2006.

\bibitem{marinetto_quantification_2016}
E.~Marinetto, M.~Brehler, A.~Sisniega, Q.~Cao, J.~W. Stayman, J.~Yorkston,
  J.~H. Siewerdsen, and W.~Zbijewski, ``Quantification of bone
  microarchitecture in ultrahigh resolution extremities conebeam {{CT}} with a
  {{CMOS}} detector and compensation of patient motion,'' in \emph{Computer
  {{Assisted Radiology}} 30th {{International Congress}} and {{Exhibition}}},
  Heidelberg, Germany, Jun. 2016, pp. S20--S21.

\bibitem{xu_technical_2016}
J.~Xu, A.~Sisniega, W.~Zbijewski, H.~Dang, J.~W. Stayman, M.~Mow, X.~Wang,
  D.~H. Foos, V.~E. Koliatsos, N.~Aygun, and J.~H. Siewerdsen,
  ``\BIBforeignlanguage{en}{Technical assessment of a prototype cone-beam
  {{CT}} system for imaging of acute intracranial hemorrhage},''
  \emph{\BIBforeignlanguage{en}{Medical Physics}}, vol.~43, no.~10, pp.
  5745--5757, Oct. 2016.

\bibitem{TilleyPenalizedLikelihoodReconstructionHighFidelity2017}
S.~Tilley, M.~Jacobson, Q.~Cao, M.~Brehler, A.~Sisniega, W.~Zbijewski, and
  J.~W. Stayman, ``Penalized-{{Likelihood Reconstruction}} with
  {{High}}-{{Fidelity Measurement Models}} for {{High}}-{{Resolution
  Cone}}-{{Beam Imaging}},'' \emph{IEEE Transactions on Medical Imaging},
  vol.~PP, no.~99, pp. 1--1, 2017.

\bibitem{AntonukEmpiricalinvestigationsignal1997}
L.~E. Antonuk, Y.~El-Mohri, J.~H. Siewerdsen, J.~Yorkston, W.~Huang, V.~E.
  Scarpine, and R.~A. Street, ``\BIBforeignlanguage{en}{Empirical investigation
  of the signal performance of a high-resolution, indirect detection, active
  matrix flat-panel imager ({{AMFPI}}) for fluoroscopic and radiographic
  operation},'' \emph{\BIBforeignlanguage{en}{Medical Physics}}, vol.~24,
  no.~1, pp. 51--70, Jan. 1997.

\bibitem{SiewerdsenEmpiricaltheoreticalinvestigation1997}
J.~H. Siewerdsen, L.~E. Antonuk, Y.~{el-Mohri}, J.~Yorkston, W.~Huang, J.~M.
  Boudry, and I.~A. Cunningham, ``\BIBforeignlanguage{eng}{Empirical and
  theoretical investigation of the noise performance of indirect detection,
  active matrix flat-panel imagers ({{AMFPIs}}) for diagnostic radiology},''
  \emph{\BIBforeignlanguage{eng}{Medical Physics}}, vol.~24, no.~1, pp. 71--89,
  Jan. 1997.

\bibitem{Starmannonlinearlagcorrection2012}
J.~Starman, J.~Star-Lack, G.~Virshup, E.~Shapiro, and R.~Fahrig, ``A nonlinear
  lag correction algorithm for a-{{Si}} flat-panel x-ray detectors,''
  \emph{Medical Physics}, vol.~39, no.~10, pp. 6035--6047, Oct. 2012.

\bibitem{StarmanInvestigationoptimallinear2011}
------, ``\BIBforeignlanguage{eng}{Investigation into the optimal linear
  time-invariant lag correction for radar artifact removal},''
  \emph{\BIBforeignlanguage{eng}{Medical Physics}}, vol.~38, no.~5, pp.
  2398--2411, May 2011.

\bibitem{NowakTimedelayedsummationmeans2012}
T.~Nowak, M.~Hupfer, F.~Althoff, R.~Brauweiler, F.~Eisa, C.~Steiding, and W.~A.
  Kalender, ``\BIBforeignlanguage{en}{Time-delayed summation as a means of
  improving resolution on fast rotating computed tomography systems},''
  \emph{\BIBforeignlanguage{en}{Medical Physics}}, vol.~39, no.~4, pp.
  2249--2260, Apr. 2012.

\bibitem{cant_modeling_2015}
J.~Cant, W.~J. Palenstijn, G.~Behiels, and J.~Sijbers,
  ``\BIBforeignlanguage{eng}{Modeling blurring effects due to continuous gantry
  rotation: {{Application}} to region of interest tomography},''
  \emph{\BIBforeignlanguage{eng}{Medical Physics}}, vol.~42, no.~5, pp.
  2709--2717, May 2015.

\bibitem{Nesterov2005}
Y.~Nesterov, ``Smooth minimization of non-smooth functions,''
  \emph{Mathematical Programming Journal, Series A}, vol. 103, pp. 127--152,
  2005.

\bibitem{Kim2015}
D.~Kim, S.~Ramani, and J.~A. Fessler, ``Combining {{Ordered Subsets}} and
  {{Momentum}} for {{Accelerated X}}-{{Ray CT Image Reconstruction}},''
  \emph{IEEE Transactions on Medical Imaging}, vol.~34, no.~1, pp. 167--178,
  2015.

\bibitem{NuytsModellingphysicsiterative2013}
J.~Nuyts, B.~De~Man, J.~A. Fessler, W.~Zbijewski, and F.~J. Beekman,
  ``Modelling the physics in iterative reconstruction for transmission computed
  tomography,'' \emph{Physics in medicine and biology}, vol.~58, no.~12, pp.
  R63--R96, Jun. 2013.

\bibitem{Long2010}
Y.~Long, J.~A. Fessler, and J.~M. Balter, ``{{3D}} forward and back-projection
  for {{X}}-ray {{CT}} using separable footprints.'' \emph{IEEE Transactions on
  Medical Imaging}, vol.~29, no.~11, pp. 1839--50, Nov. 2010.

\bibitem{0031-9155-60-4-1415}
A.~Sisniega, W.~Zbijewski, J.~Xu, H.~Dang, J.~W. Stayman, J.~Yorkston,
  N.~Aygun, V.~Koliatsos, and J.~H. Siewerdsen, ``High-fidelity artifact
  correction for cone-beam {{CT}} imaging of the brain,'' \emph{Physics in
  Medicine and Biology}, vol.~60, no.~4, p. 1415, 2015.

\bibitem{huber_robust_statistics}
P.~J. Huber, \emph{Robust Statistics}.\hskip 1em plus 0.5em minus 0.4em\relax
  New York: {Wiley}, 1981.

\end{thebibliography}

\end{document}